\documentclass{aa}  
\usepackage{graphics}
\usepackage[authoryear]{natbib}
\bibpunct[]{(}{)}{;}{a}{}{,}

\begin{document}

\def\gtapp
{\mathrel{\hbox{\raise0.3ex\hbox{$>$}\kern-0.8em\lower0.8ex\hbox{$\sim$}}}}
\def\ltapp
{\mathrel{\hbox{\raise0.3ex\hbox{$<$}\kern-0.75em\lower0.8ex\hbox{$\sim$}}}}

\title{Extended mid-infrared emission from VV\,114: probing the birth of a
ULIRG\thanks{Based on observations with the ISO satellite, an ESA
project with instruments funded by ESA Member States (especially the
PI countries: France, Germany, the Netherlands and the United Kingdom)
and with the participation of ISAS and NASA.}}
  
\author{E.~Le~Floc'h \inst{1}  
\and V.~Charmandaris \inst{2} 
\and O.~Laurent \inst{1,3}  
\and I.F.~Mirabel \inst{1,4}  
\and P.~Gallais \inst{1}  
\and M.~Sauvage \inst{1}  
\and L.~Vigroux \inst{1}  
\and C.~Cesarsky \inst{5}  
}
  
\institute{CEA/DSM/DAPNIA Service d'Astrophysique, F-91191 Gif-sur-Yvette,  
France  
\and  
Cornell University, Astronomy Department, Ithaca, NY 14853, USA  
\and  
Max Planck Institut f\"ur extraterrestrische Physik, Postfach 1312, 85741  
Garching, Germany  
\and   
Instituto de Astronom\'\i a y F\'\i sica del Espacio, cc 67, suc 28.   
1428 Buenos Aires, Argentina
\and   
European Southern Observatory, Karl-Schwarzschild-Str, D-85748   
Garching bei M\"unchen, Germany
} 
  
\titlerunning{Mid-infrared observations of VV\,114 with ISOCAM}
\authorrunning{Le Floc'h et al.}
  
\offprints{E. Le Floc'h (elefloch@cea.fr)}  
  
\date{Received March 1, 2002 / Accepted May 23, 2002}  
  
\abstract{ 
We present our 5--16\,$\mu$m spectro-imaging observations of
\object{VV\,114}, an infrared luminous early-stage merger of two
galaxies VV\,114E and VV\,114W, taken with the ISOCAM camera on-board
the {\it Infrared Space Observatory}. We find that only 40\% of the
mid-infrared (MIR) flux is associated with a compact nuclear region of
VV\,114E, while the rest of the emission originates from a rather
diffuse component extended over several kpc in the regions between
VV\,114E and VV\,114W. This is in stark contrast with the very compact
MIR starbursts usually seen in luminous and ultraluminous infrared
galaxies. A secondary peak of MIR emission is associated with an
extra-nuclear star forming region of VV\,114W which displays the
largest H$\alpha$ equivalent width in the whole system. Comparing our
data with the distribution of the molecular gas and cold dust, as well
as with radio observations, it becomes evident that the conversion of
molecular gas into stars can be triggered over large areas at the very
first stages of an interaction. This extended star formation along
with the extreme nuclear starburst observed in VV\,114E can easily
lead to the heating of dust grains found in the tidally disturbed
disks of the progenitor galaxies and subsequently traced via their MIR
emission. The presence of a very strong continuum at the
5--6.5\,$\mu$m range in the spectrum of VV\,114E indicates that an
enshrouded active galactic nucleus (AGN) may contribute to $\sim$ 40\%
of its MIR flux. We finally note that the relative variations in the
UV to radio spectral properties between VV\,114E and VV\,114W provide
evidence that the extinction-corrected star formation rate of similar
objects at high z, such as those detected in optical deep surveys, can
not be accurately derived from their {\it rest-frame} UV properties.
}  

\maketitle
  
\section{Introduction}  

The first all-sky survey at far-infrared (FIR) wavelengths performed
by IRAS in 1983 led to the discovery of a large number of galaxies
with L$_{\rm IR}\geq10^{11}$\,L$_{\sun}$, the so-called luminous
infrared galaxies \citep[see][]{rieke,soifer89}. Ever since, extensive
work has been devoted to better understand the nature of these
objects which emit the bulk of their energy in the 8--1000\,$\mu$m
domain, and
it is now widely accepted that the brightest of these sources are
formed during the collisions of massive gas-rich late type galaxies
\citep[see][ and references therein]{sanders96}. In such galactic
encounters, as the tidal interactions strip angular momentum from the
interstellar matter, large concentrations of molecular gas are indeed
driven by viscous accretion to the few central 100\,pc of the merging
galaxies \citep{solomon,combes}, feeding at high rates powerful
circumnuclear starbursts and/or supermassive black holes
\citep{sanders88}.  The huge amounts of dust accumulated into the
central regions lead to a significant absorption of the UV and optical
radiation which in turn is thermally reprocessed at longer
wavelengths, and is thus responsible for the luminous phase in the
infrared.  Several key problems though, such as the time evolution and
duration of the infrared burst during the merger event, as well as the
identification of the dominant sources heating the dust (stars, AGN)
are still not fully answered.  Observations of colliding galaxies at
different merging stages are therefore of a prime importance to better
understand how the AGN and starburst phenomena are related
\citep[see][]{mihos96}.

The interest in the study of these objects was further increased over
the past few years when it became evident that they also played a
critical role in galaxy evolution {\it over cosmological look-back
times}. Luminous infrared galaxies of the local Universe are indeed
excellent analogs to the high-redshift dusty objects recently
discovered in deep mid-infrared and sub-millimeter surveys
\citep[e.g.][]{elbaz,hughes}. They are thus considered as the
low-z counterparts of a population of dust-enshrouded sources which
were formed at an epoch when interactions and mergers among galaxies
were much more frequent than today \citep{lefevre}.  Consequently,
investigating the properties of colliding galaxies and the formation
mechanisms of sources luminous in the infrared (IR) at low redshift
can provide valuable insights on the evolution of galaxies in the
distant Universe.

VV\,114 (= Arp\,236 = IC\,1623), located at a distance of
80\,Mpc\footnote{Throughout the paper, we adopt
H$_0$\,=\,75\,km\,s$^{-1}$\,Mpc$^{-1}$ and q$_o$\,=\,0.5 to facilitate
comparison with earlier work.}(1{\arcsec}=390\,pc), is one of such
interacting systems undergoing vigorous starburst activity.  Its
infrared luminosity is L$_{\rm IR}$ = $10^{11.6}$ L$_{\sun}$
\citep{soifer87}, making it a LIRG\footnote{
We define LIRGs as galaxies with 10$^{11}$\,L$_{\sun}\leq$\,L$_{\rm 
IR}$$\leq$\,10$^{12}$\,L$_{\sun}$, and ULIRGs as ultraluminous
sources with L$_{\rm IR}$$\geq$10$^{12}$\,L$_{\sun}$. We keep the more
general expression of ``luminous infrared galaxies'' for those
with L$_{\rm IR}$$\geq$10$^{11}$\,L$_{\sun}$.}
and one of the brightest
objects of the IRAS Bright Galaxy Sample.  It appears to be an
early-stage merger of two galaxies (see Figure~\ref{fig:iso_view})
which are aligned east-west with a projected nuclear separation of
$\sim$ 6\,kpc and therefore referred in the literature as VV\,114E and
VV\,114W \citep{knop}. The global properties of the system are
summarized in Table~\ref{tab:properties}.  At optical wavelengths,
VV\,114 shows a highly disturbed morphology with very faint tidal
tails extending over 25\,kpc from the center \citep{arp}.  The western
component, VV\,114\,W, is more extended than the eastern one, and
dominates the emission in the visible. Three peaks of intensity are
clearly visible in the V band and their optical spectrum indicate an
emission typical of \ion{H}{ii} regions with an average visual
extinction $A_v = 1.7$~mag \citep{knop}.  While the brightest peak to
the North (knot 1) probably coincides with the dynamical center of
VV\,114W \citep{doyon}, the maximum of star formation traced by the
H$\alpha$ line is displaced and it occurs in the secondary peak
(knot~2) which has an H$\alpha$ equivalent width six times larger than
knot~1, and is located 5{\arcsec} to the southeast in the overlap
region of the two galactic disks
\citep{knop}.

The near-infrared (NIR) morphology and spectral properties of VV\,114
present a remarkable contrast with what is observed in the
visible. Medium resolution NIR spectra of both components display
strong recombination lines of hydrogen and helium as well as deep CO
bands at 2.3\,$\mu$m, indicating vigorous and widespread star
formation in both galaxies \citep{doyon}. However, whereas the JHK
colors of VV\,114W are still consistent with the presence of numerous
\ion{H}{ii} regions, the continuum emission from VV\,114E which is nearly
invisible at 0.44\,$\mu$m (B band), becomes brighter as the wavelength
increases.  Its red colors indicate large concentrations of dust
\citep[$A_v \sim$ 4\,mag,][]{doyon} obscuring a luminous source which is
unveiled at 2\,$\mu$m and dominates the whole emission of the system
in the near-infrared \citep{knop}.  High resolution K-band images of
this eastern core taken with HST/NICMOS reveal a double-nucleus
\citep[see Fig.~2 in][]{scoville} with the two near-infrared sources
separated by 1.6{\arcsec} ($\sim$ 630~pc). The NICMOS data clearly
show numerous compact \ion{H}{ii} regions and star clusters distributed
along a spiral arm in the overlap of the merger, which suggests an
enhancement of the starburst activity in this area \citep{scoville},
in agreement with the H$\alpha$ observations \citep{knop}.

The presence of a powerful starburst in VV\,114 is also suggested
by its huge amount of molecular gas \citep[5.1$\times$10$^{10}$
M$_{\sun}$, see][]{yun}, which makes this merger one of the most gas-rich
systems in the local Universe.  The CO emission is distributed along a
bar-shaped structure joining VV\,114E and VV\,114W, with a peak of
intensity centered between the two merging galaxies and its irregular
kinematics suggest that the gas has not yet settled
\citep{yun}. Submillimeter observations performed with SCUBA on
the JCMT have also detected an extended dust emission over 30{\arcsec}
(12\,kpc), which correlates well with the spatial distribution of the
CO emission \citep{frayer} and peaks just 5{\arcsec} southwest of
VV\,114E. The flux observed at 450 and 850\,$\mu$m is in excess of that
expected from the IRAS data and suggests the presence of a cool
(T $\sim$ 20-25\,K) and massive ($10^8$ M$_{\sun}$) dust component
\citep{frayer}. Finally, VV\,114 also exhibits a strong radio
continuum emission, with a ratio of FIR to radio radiation typical of
that observed in the IRAS Bright Galaxy Sample \citep{condon90}. Its
1.49\,GHz emission globally follows the distribution seen in the
visible and it extends along the overlap region between the east and
western components. A local maximum of the radio emission is seen to
coincide with the brightest H$\alpha$ emission in the overlap region,
while VV\,114E displays a double-peak radio morphology similar to that
observed at 2.2\,$\mu$m \citep{condon90,condon91}. The spatial
coincidence of these two sources at NIR and radio wavelengths combined
with the similarity of the two nuclei in terms of the NIR/radio flux
ratio and the CO kinematics suggest that VV\,114E itself may have
experienced an earlier merger.

\begin{table}[htbp]
\caption{Global properties of VV\,114.}
\begin{tabular}{p{5.5cm}r}
\hline 
\hline 
Parameter & Value \\
\hline 
$\alpha$\,(J2000.0) VV\,114E~$^a$ 	& 01$^h$07$^m$47$^s_.$5 \\
$\delta$\,(J2000.0) VV\,114E~$^a$  	& -17$\degr$30$'$25$''$ \\ 
Distance$\,^b$				& 80~Mpc \\
L$_{\rm IR}\,^c$			& 10$^{11.6}$\,L$_{\sun}$ \\
$f_{25 \mu m}/f_{60 \mu m}\,^c$		& 0.18 \\
M$_{\rm H_2}\,^d$			& $5.1\times10^{10}$\,M$_{\sun}$\\
L$_{\rm IR}$/M$_{\rm H_2}$\,(L$_{\sun}$/M$_{\sun}$)$\,^d$ & 7.8 \\
\hline
\hline
\end{tabular} \\
\vspace{.015cm} \\
$^a$  Coordinates refer to the northeastern nucleus of VV\,114E. \\
$^b$  We use H$_0$\,=\,75\,km\,s$^{-1}$\,Mpc$^{-1}$, q$_o$\,=\,0.5 \\
$^c$  From \citet{soifer87}. \\
$^d$ Based on\citet{yun}. \\
\label{tab:properties}
\end{table}

Since observations in the mid-infrared (MIR) are less affected by
absorption than in the visible \citep[A$_{15\,\mu m}$
$\sim$\,A$_{V}$/70, see][]{mathis}, they provide a powerful tool to
probe deeper into the dust-enshrouded regions of galaxies, and allow
one to trace the reprocessed emission of dust heated by UV
radiation. Recent high resolution MIR images of VV\,114 obtained
at Keck by \citet{soifer01} led to a significant improvement in our
understanding of that system -- discussed in detail in the following
sections -- despite the fact that ground-based observations generally
suffer from the low transparency of the atmosphere at these
wavelengths.  We thus used the high sensitivity and the good spatial
and spectral resolution of the ISOCAM camera to gain a new insight in
the mid-infrared properties of VV\,114.  After describing the ISO
observations and data reduction in section~\ref{sec:red}, we present
our analysis on the distribution of the MIR emission in
section~\ref{sec:morpho} and the spectral properties of VV\,114 in
section~\ref{sec:spectra}. We discuss our results in
section~\ref{sec:discuss} and summarize our conclusions in
section~\ref{sec:conclusions}.

\section{Observations, reduction and analysis}  
\label{sec:red}  
  
\subsection{The data}

\begin{figure}  
\resizebox{\hsize}{!}{\includegraphics{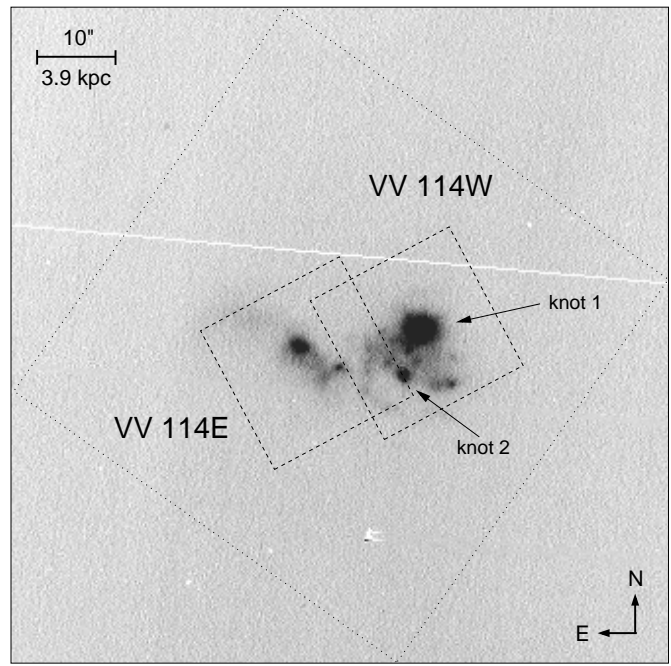}}
  \caption{An I-band image of VV\,114 obtained from the CFHT
  archive. The two interacting galaxies, as well as the regions knot 1
  and 2 of \citet{knop} are marked. The field of view of the ISOCAM
  15\,$\mu$m LW3 image is shown with the dotted square, while the two
  NICMOS pointings by \citet{scoville} which were used to construct the
  mosaic presented in Figure~\ref{fig:mosaic} are indicated with the
  dashed line. The scale and orientation of the frame are also
  displayed.}
\label{fig:iso_view}
\end{figure}

VV\,114 was observed with the ISOCAM camera \citep{cesarsky}, a
32$\times$32 pixel array on-board the Infrared Space Observatory
\citep{kessler}. The data were obtained on 6 January 1998 in
low-resolution spectro-imaging mode with the Continuously Variable
Filter (CVF) covering the full 5.1--16.3\,$\mu$m wavelength range with
a spectral resolution between 30 and 40. We used a 2.1\,sec
integration per CVF frame spending a total of 2.5\,hrs on source. The
pixel size was 1.5{\arcsec} creating a field of view of
48{\arcsec}$\times$48{\arcsec}, and the full-width at half-maximum
(FWHM) of the point spread function (PSF) varied between 4{\arcsec}
and 5{\arcsec}. In order to better study the low surface brightness of
the MIR emission with a better signal to noise ratio than what can be
achieved with the CVF, we also retrieved from the ISO
archive\footnote{The ISO archive is available at
www.iso.vilspa.esa.es/ida/} a 15$\mu$m ISOCAM image of the galaxy
obtained using the wide LW3 (12--18\,$\mu$m) filter. This 3$\times$3
raster map was also obtained using a pixel size of 1.5{\arcsec}, a
6{\arcsec} step in both axes, and as a result it covered a
1{\arcmin}$\times$1{\arcmin} field encompassing the whole optical
extent of the galaxy (see Figure~\ref{fig:iso_view}).

The data were reduced using the CAM Interactive Analysis
(CIA\footnote{CIA is a joint development by the ESA astrophysics
division and the ISOCAM consortium.}) software following the standard
procedure described in detail by \citet{starck}.  In brief we
performed: 1) the subtraction of the dark current, taking into account
the observing time parameters, 2) the cosmic-ray removal by applying a
wavelet transform method, and 3) the correction of detector memory
effects using the Fouks-Schubert's method \citep{coulais}.  The sky
was subtracted using the emission-free regions of the detector outside
the galaxy. The differential pixel-to-pixel response of the array was
corrected using flat-field images taken as part of the ISOCAM
calibration. Finally, the effects of jitter were also corrected. These
effects result from the combination of the satellite tracking motion
($\sim$1{\arcsec} in amplitude), the continuous translation of the
source on the detector as a function of the observed wavelength, which
is an intrinsic feature of the ISOCAM CVF optics, and the
discontinuous shift of the target when changing from one CVF sector to
another. To perform the jitter correction, all frames were resampled
with a smaller pixel size and aligned with one another using the
brightest peak of MIR emission observed in each frame. 

\subsection{Astrometry}
\label{subsec:astro}

Due to the absolute pointing uncertainty of the satellite, the rather
small field of view of our MIR images and the lack of bright stars
in the field, special care had to be taken to define the proper
astrometry of our observations.  For this purpose, we used the
diffraction-limited images of VV\,114 recently obtained by
\citet{soifer01} with the MIRLIN mid-infrared camera \citep{ressler}
on the Keck\,II telescope. These high resolution data indicate that
1)~the maximum of intensity detected at 12.5\,$\mu$m originates
from the core of VV\,114E, 2)~the double-nucleus morphology of
VV\,114E observed at NIR and radio wavelengths is also present in the
mid-infrared, and 3)~both nuclei (referred as VV\,114E$_{NE}$ and
VV\,114E$_{SW}$) have approximately the same flux at 12.5\,$\mu$m
\citep[see Fig.\,1a of ][]{soifer01}. Given their small angular
separation of 1.6{\arcsec}, these two sources can not be directly
resolved in our data and therefore we set the peak observed in the
ISOCAM images between the two (see Figure~\ref{fig:mosaic}b).  
We note that \citet{soifer01} reported sub-arcsec
scale discrepancies regarding the spatial coincidence of these two
nuclei as seen at NIR, MIR and radio wavelengths. These potential
misalignments though are much smaller than the FWHM of the ISOCAM PSF
and do not significantly affect our astrometry and analysis. We did
not make any modification to the roll angle of our maps since it is
unlikely to suffer from systematic errors given its observed
performance over the ISO mission. In fact no uncertainties in the
rotation angle were ever reported in ISOCAM deep surveys where they
would have been easily observed due to the large number of detections
over large spatial scales.  We estimate that our final absolute
astrometry is accurate to $\sim$0.5{\arcsec}.

\subsection{Photometry}
\label{sec:photo}

The 15\,$\mu$m image of VV\,114 (see Figures~\ref{fig:mosaic}a and
\ref{fig:mosaic}b), taken with the LW3 filter, shows that the MIR
emission can be roughly decomposed into a bright peak at the nuclear
region of VV\,114E and a more extended component towards VV\,114W.
This diffuse emission, which was not detected in the MIRLIN
images of \citet{soifer01}, has an average surface brightness of
1.3\,mJy\,arcsec$^{-2}$. It is not particularly pronounced at the
optical center of the western galaxy, but it presents a peak of
2.2\,mJy\,arcsec$^{-2}$ along the interacting region between the two
galaxies \citep[marked as knot~2 by][, see also
Figure~\ref{fig:iso_view}]{knop}. Even though the authors set a
slightly higher upper limit of 1\,mJy\,arcsec$^{-2}$ for it, this
difference can be understood as due to the higher sensitivity and
larger pixel size of our ISOCAM maps.

In the analysis of our data on VV\,114E, a substantial effort was
devoted to determine whether the MIR ISOCAM profile can be attributed
only to the double nucleus at its center, or if emission from an
additional component associated with an extended circumnuclear region
is also present. For this purpose, we modeled the observed profile
with a combination of two point sources simulating the two nuclei of
VV\,114E observed by \citet{soifer01}. As a basis for our model we
used a series of theoretical PSFs calculated for the specific
configuration of ISOCAM and for various offsets of the PSF maximum
relative to the pixel centers on the grid of the detector
\citep{okumura}.  Taking into account 1)~the pixel scale, 2)~the
orientation angle of our maps, and 3)~the position angle of the two
nuclei as well as their angular separation of 1.6{\arcsec}, we thus
obtained a collection of simulated profiles for each wavelength of the
CVF.  We then applied an iterative fitting method using a $\chi^2$
test to select the best simulation available.  Since the relative
contribution of VV\,114E$_{NE}$ and VV\,114E$_{SW}$ is not accurately
known over the entire wavelength range of the CVF, we assumed that the
two sources contribute the same level of emission throughout the
5--16\,$\mu$m range, and compared the result with the extreme
situation where all emission originates from VV\,114E$_{SW}$. In both
cases we found that for the best simulated profile, the fit can not
account for all emission from VV\,114E. The observed MIR excess is
present over the whole 5--16\,$\mu$m range and can be attributed to
the contribution from an extended component surrounding the two nuclei
over an area of a few 100\,pc. Before attempting to quantitatively
separate this extended emission from the one originating from the two
nuclei, we wanted to ensure that the former is not artificially
produced by a wrong correction of the jitter effect in our data. To
achieve this, we used the recorded telemetry information of the
telescope, and simulated how the profile of a single PSF would widen
if no jitter correction was applied. We found that the effect is
noticeable only at short wavelengths, and that the error on the
measured flux is negligible (less than 1\%) provided that an aperture
of at least 3$\times$3 pixel is used.  We therefore chose to measure
the intensity from the double nucleus by integrating the flux over a
fixed aperture of 3$\times$3 pixels
(4.5{\arcsec}$\times$4.5{\arcsec}). Using the best simulated profiles
obtained with the $\chi^2$ method described earlier, we found that the
excess due to the extended emission around the nucleus of VV\,114E in
the aforementioned aperture is very small at long wavelengths
($\sim$10\% beyond 12\,$\mu$m), but varies between 30 and 60\% at
5\,$\mu$m depending on the assumed relative contribution of the two
nuclei.  Finally, we restituted the spectral energy distribution of the
double nucleus by applying an aperture correction to take account of
the spatial extent of the PSF.  This correction was computed assuming
a 4.5\arcsec$\times$4.5\arcsec aperture and {\it a single point
source} at the center of VV\,114E. Using our simulations described
above, it can be shown indeed that the fraction of flux found outside
of the above apertures is only weakly sensitive to the fact that
VV\,114E has two nuclei separated by $\sim$1.6\arcsec.

We also compared our observed spectral energy distribution (SED) with
the broad-band photometry of the central region of VV\,114E as derived
by \citet{soifer01}.  This was performed by integrating our spectrum
over the appropriate wavelength ranges, using the profile of the
MIRLIN filters.  Our ``MIRLIN equivalent'' broad band fluxes in all
five bands are in good agreement with those observed by
\citet{soifer01} with an averaged uncertainty of 10\%.  This
consistency indicates that our assumptions and analysis resulted in a
reliable MIR spectrum from the nucleus of VV\,114E. Furthermore, our
data indicate that VV\,114E accounts for only 40\% of the total MIR
flux of VV\,114 detected by ISOCAM. This also follows the observation
by \citet{soifer01} that a substantial fraction of the IRAS 12\,$\mu$m
flux\footnote{See Appendix A for a discussion on the IRAS
measurements.} is not detected in their data, and may originate from a
more diffuse component, which is in fact detected by
ISOCAM. Integrating over the full spatial extent of VV\,114E we find
that the total extended emission in VV\,114E accounts for 35\% of the
flux beyond 12\,$\mu$m, and reaches 50 to 70\% at shorter wavelengths.
To examine the spatial distribution of that diffuse component, we used
the high signal to noise 15$\mu$m image and removed the contribution
of the double nucleus by subtracting the best simulations selected
with our $\chi^2$ test as mentioned above. This revealed a secondary
peak of emission 3{\arcsec} west of the double nucleus of VV\,114E,
which is totally inconspicuous in the images dominated by the strong
flux emitted by the core of the eastern galaxy. The maximum of its
surface brightness is $\sim$2\,mJy\,arcsec$^{-2}$, and interestingly
it coincides spatially with the peak of CO emission \citep{yun}.
Yet, the reality of this feature remains somewhat uncertain since it
depends on the simulated profile used for the removal of the double
nucleus. Moreover, it does not have any counterpart at NIR or radio
wavelengths, and may likely result from an artifact in our PSF
subtraction.

\begin{table*}[htbp]
\caption{ISOCAM broad-band photometry of VV\,114.}
\begin{tabular}{p{3.5cm}ccccccc}
\hline 
\hline 
\vspace{.1cm} 
 & \multicolumn{7}{c}{ISOCAM flux in broad-band filters (mJy)} \\ 
\vspace{.01cm}  
 & LW2 & LW3 & LW4 & LW6 & LW7 & LW9 & LW10$^a$ \\
Source & (5-8.5\,$\mu$m) &  (12-18\,$\mu$m) & (5.5-6.5\,$\mu$m) & 
(7-8.5\,$\mu$m) & (8.5-10.7\,$\mu$m) & 
(14-16\,$\mu$m) & (8-15\,$\mu$m)   \\
\hline 
VV\,114E (nucleus)   & 136 & 351 &  96 & 242 & 105 & 379 & 198 \\
VV\,114W (knot 2)    &  28 &  88 &  13 &  62 &  35 &  85 &  58 \\
VV\,114 (diffuse)    & 256 & 241 & 111 & 296 & 180 & 186 & 264 \\
\hline
VV\,114 (total)      & 420 & 680 & 220 & 600 & 320 & 650 & 520 \\
\hline
\hline
\end{tabular}
\vspace{.1cm} \\
$^a$  equivalent to IRAS 12\,$\mu$m filter. \\
Note: Fluxes in the LW3 filter were measured directly in the LW3
broad-band image retrieved from the ISO archive. Equivalent fluxes
in the other filters were obtained from the CVF data,
those of VV\,114E (nucleus) and VV\,114W (knot~2) 
being derived from their MIR SED respectively presented on 
Figure~\ref{fig:spec_vv114w} and Figure~\ref{fig:spec_vv114e}. 
The diffuse emission was deduced after subtracting
the contributions of VV\,114E (nucleus) and knot~2 to the total 
fluxes measured over the full spatial extent
of the merger.

\label{tab:fluxes} 
\end{table*}

Finally using our MIR spectrum, we calculated the equivalent flux
densities corresponding to the typical ISOCAM broad-band filters for
several components of VV\,114. 
Taking account of the filter transmission curves, we integrated
the SED from the CVF to obtain energy fluxes in W\,m$^{-2}$. The
conversion to derive the flux densities in mJy 
was performed assuming an intrinsic
spectral shape $f_{\nu}$ = cte. The total intensity for the two
galaxies was obtained by integrating the MIR emission over the full
spatial extent of the galactic disks as seen in our data. We note that
within 1\,$\sigma$ of our photometric uncertainty, the flux measured
by the LW10 filter, an equivalent to the 12\,$\mu$m IRAS filter,
accounts for nearly 100\% of the IRAS flux, the exact value of which
we discussed in Appendix~A. The flux from the central region of
VV\,114E was derived as explained earlier, and the one from the peak
of the extended emission located southeast of VV\,114W was measured
over a region of 5{\arcsec} in diameter centered on knot~2 ($\alpha =
01^h07^m46.7^s, \delta = -17\degr30'27''$, J2000). Our results are
presented in Table~\ref{tab:fluxes}. The 1$\sigma$--rms noise in our
images created from the CVF by integrating over the 5--8.5\,$\mu$m
(equivalent LW2 filter) and 12--16\,$\mu$m wavelength range are 150
and 230\,$\mu$Jy\,arcsec$^{-2}$ respectively.  Clearly, these images
are not as deep as the one obtained directly with the broad-band LW3
filter in which we measured a noise level of
70\,$\mu$Jy\,arcsec$^{-2}$. This was to be expected since
the bandpass of the CVF is much narrower than the
width of the  broad-band filters.
The relative photometric uncertainty on the
MIR SED varies from 10$\%$ at 5\,$\mu$m up to 15$\%$ at
16\,$\mu$m. The major sources of errors are due to the non-perfect
correction of the detector memory effect. Based on our experience with
ISOCAM data, we estimate that an uncertainty of 25\,$\%$ for the
absolute photometry constitutes a conservative upper limit, 
which is typical for well detected extended
sources.

\subsection{Mapping of the UIB emission}
\label{subsec:uib}

As we will discuss in more detail in section~\ref{sec:spectra}, it is
widely accepted that the various Unidentified Infrared Bands (UIBs)
detected in MIR spectra of galaxies are produced in photo-dissociation
regions (PDRs) and consequently are directly associated with the
regions of star formation in galaxies
\citep[see][]{rigopoulou,laurent,helou}. Their spatial distribution
could then be used to depict more accurately the regions undergoing
star forming activity even in areas where extinction from dust is
high. We thus extracted the UIB emission from our CVF spectral cube
and produced images of their spatial distribution across VV\,114,
following the same method as in \citet{lefloch}. We identified the
6.2, 7.7 and 11.3\,$\mu$m bands in all observed positions (pixels) on
the galaxy and subtracted the underlying continuum emission (see
Figure~\ref{fig:spec_vv114w} for an illustration of the method). Since
on galactic scales the relative strength of the different UIBs does
not vary substantially, we co-added the corresponding images to
produce a single higher signal to noise map of the integrated UIB
emission.  We decided not to include the 8.6\,$\mu$m feature which is
highly sensitive to the strength of the 9.7\,$\mu$m silicate
absorption band. The 12.7\,$\mu$m feature was also excluded since it
is contaminated by the [NeII] line at 12.8\,$\mu$m (see
section~\ref{sec:HII}).

\section{Results}
\subsection{Spatial distribution of the MIR emission}
\label{sec:morpho}

\begin{figure*}[!ht]  
\resizebox{\hsize}{!}{\includegraphics{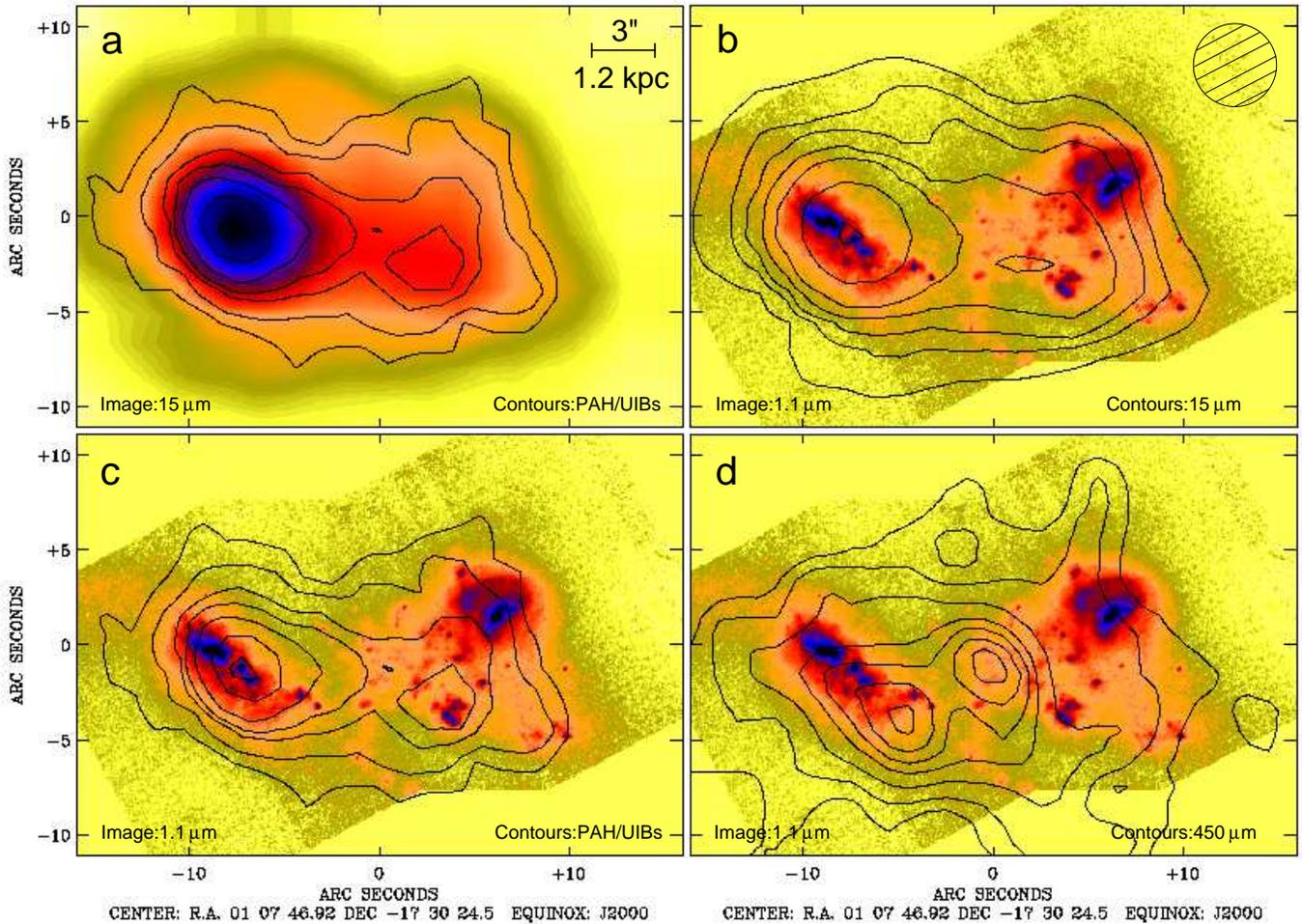}}  
  \caption{a) The ISOCAM 15\,$\mu$m image of VV\,114 taken with LW3
  (12--18\,$\mu$m) filter with an overlay of the integrated UIB
  emission at 6.2, 7.7 and 11.3\,$\mu$m (see
  section~\ref{subsec:uib}). 
  The image scale is displayed with a bar of 1.2\,kpc (3\,{\arcsec})
  at the top-right corner. b) The HST/NICMOS J-band image of VV\,114
  from \citet{scoville} with the overlay of the LW3 15\,$\mu$m image
  presented in a). The contour levels are 0.45, 0.65, 0.90, 1.35,
  2.10, 4.45 and 8.45~mJy\,arcsec$^{-2}$.  
  We have also indicated the FWHM of the ISOCAM PSF at 12\,$\mu$m.
  Note that the warm dust
  emission is much more extended than the nuclei of the galaxies. 
  c) The same J-band image presented in b) but overlayed with the UIB
  contours of a). Note how the secondary UIB peak is not found in
  VV\,114W but instead it coincides with knot 2 of \citet{knop}. d) An
  overlay of the SCUBA 450\,$\mu$m emission detected by \citet{frayer}
  onto the NICMOS image presented in b) and c). The contours are 150,
  225, 338, 450, 525, 600 and 675~mJy\,beam$^{-1}$ while the beam size
  is $\sim 6''$. We observe that, similar to the CO emission, most of
  the cold dust is found in the area between the galaxies suggesting
  that there is a considerable reservoir of gas to fuel the star
  formation activity in the system.}
\label{fig:mosaic}
\end{figure*}

In Figure~\ref{fig:mosaic}a, we overlay in contours this UIB map on a
15\,$\mu$m image of VV\,114 obtained with the LW3 filter while in
Figures~\ref{fig:mosaic}b and c, we present a mosaic of the two
1.1\,$\mu$m J-band images of VV\,114 taken with
HST/NICMOS\footnote{Individual frames are available at the NASA
Extragalactic Database hosted in the web server of IPAC, see also
\citet{scoville}.} superimposed with the contours of the 15\,$\mu$m
and UIB emission respectively. One can see that the UIB emission seems
to be more closely restricted to the active regions of the galaxy than
the widespread distribution of hot dust seen in the 15\,$\mu$m image.
This apparent lack of UIB features in the outskirts of VV\,114 is
most probably due to a lower sensitivity of our UIB maps.

The power of infrared to probe deeper into the embedded regions of
galaxies is particularly enlightening in the case of VV\,114E, the
activity of which becomes apparent in the thermal MIR even though it
is rather inconspicuous in the optical. We clearly see that this
source dominates the MIR emission of VV\,114 (see
Figure~\ref{fig:mosaic}a), and roughly accounts for 40\% of the total
flux observed. The fact that VV\,114E has an LW3 to K-band ratio of
$\sim$ 16 which is larger than the 1--10 range of values found in late
type galaxies \citep{boselli}, suggests a higher than average hot dust
content compared to its stellar mass. Strong UIB features are also
seen in its spectrum (see Figure~\ref{fig:mosaic}c) further confirming
that an important fraction of the flux coming from the central regions
of the eastern galaxy originates from a powerful circumnuclear
starburst (see also section~\ref{sec:nuc}). Our findings support
similar conclusions derived by \citet{doyon} based on NIR
spectroscopic observations. Note that the intense star forming
activity in the inner core of VV\,114E had also been inferred from the
strong 8.4\,GHz radio emission
\citep{condon91}, which coincides with the bright NIR peaks
\citep{knop}, as well as the large quantity of molecular gas in the
vicinity of VV\,114E \citep{yun}.

Contrary to VV\,114E, the MIR emission from VV\,114W is weaker and
more diffuse. Surprisingly, there is no enhancement of the MIR flux
near its nucleus. Instead, the dust emission seen with ISOCAM peaks
5{\arcsec} to the southeast, close to the bright \ion{H}{ii} region
denoted as knot~2 by \citet{knop}. This secondary peak at knot 2 is
even more pronounced in our UIB map and is also visible in the
3.2\,$\mu$m image of \citet{soifer01} which sampled emission from the
weaker 3.3\,$\mu$m UIB feature. Knot 2 also coincides with a local
maximum of the radio emission at 1.49\,GHz \citep{condon90}, and has
an H$\alpha$ equivalent width of 430\,\AA, six times wider than what
is observed in the nucleus of VV\,114W, a clear evidence of massive
extranuclear star formation activity \citep{knop}. A large number of
young star clusters and compact \ion{H}{ii} regions were also detected
in the NIR throughout the same area \citep{scoville}. As a whole
though, VV\,114W has an LW3/K ratio $\sim$~11 slightly higher than
that of normal galaxies, and accounts for $\sim$ 35\% of our observed
total MIR flux.

Another striking feature in our maps is the spatial extent of the
diffuse MIR emission. This is clearly noticeable in the 15\,$\mu$m
image of VV\,114 (see Figure~\ref{fig:mosaic}b). Its distribution is
rather smooth with an elliptical shape surrounding the central regions
of the two merging galaxies over more than 10\,kpc.  This is in stark
contrast to what is observed in most LIRGs and ULIRGs which typically
harbor very compact circumnuclear starbursts responsible for nearly
80--100\% of the energy emitted in the mid-infrared
\citep[i.e.][]{soifer00, soifer01,charmandaris02b}. As was already
noted by \citet{soifer01} though, VV\,114 is an exception to this
trend since its 12\,$\mu$m and NIR curves-of-growth are quite
similar. We explore the implications of this finding in
section~\ref{sec:discuss}.

At kpc scales, the diffuse and extended emission from hot dust
particles as seen in our data closely follows the distribution of the
molecular gas detected in the CO interferometric map of
\citet{yun}. The CO emission reveals a fairly extended and massive
molecular gas complex at the center of the merger, distributed along a
bar joining the two galaxies. Two tail-like features extend from the
extremities of this bar, one to the North of VV\,114W, and the other
to the South of VV\,114E. This morphological correspondence between CO
and MIR emission in tails or spiral arms has already been observed
both in normal and in interacting galaxies and AGN
\citep{mirabel99,wilson00,lefloch} and is to be expected as the
molecular gas is the fuel of star formation which in turn heats
the dust to emit in the MIR.

Recently, submillimeter observations of VV\,114 by \citet{frayer} at
450\,$\mu$m and 850\,$\mu$m with SCUBA have also revealed a cold dust
emission centered in the overlap regions between the two interacting
components and extended over 12\,kpc. As we can see from
Figure~\ref{fig:mosaic}d, despite the large SCUBA beam (6{\arcsec} at
450\,$\mu$m) the cold dust follows the same overall distribution as
our ISOCAM maps, but the two peaks of the 450\,$\mu$m emission are
slightly displaced relative to the MIR. This can be interpreted by
temperature gradients within the dust reservoir which is heated by
both the nucleus of VV\,114E and the star-forming source at knot\,2.
The distribution of the cold dust seen at 850\,$\mu$m also correlates
well with the CO emission given the uncertainty ($\pm$ 2\,{\arcsec})
of the SCUBA map, and seems to peak in the overlap region of the
merging galaxies just to the southwest of VV\,114E.  However,
determining the location where the bolometric luminosity of VV\,114
mainly originates is not straight forward. According to the models of
the FIR emission presented in Figure~2 of \citet{frayer}, most of the
energy output in this merger is radiated between 60 and
200\,$\mu$m. We note that the emission probed by SCUBA only accounts
for less than one tenth of the total luminosity and is similar to the
contribution of the MIR emission traced by ISO at 15\,$\mu$m.  This
suggests that the 60\---200\,$\mu$m emission is probably concentrated
between the peaks of the MIR and submillimeter emission, and thus the
bolometric luminosity of the system should be more closely associated
with the eastern component of VV\,114.

\subsection{Spectral properties}
\label{sec:spectra}

\subsubsection{The Mid-infrared SED of VV\,114W}
\label{sec:HII}

\begin{figure}[!ht]  
\resizebox{\hsize}{!}{\includegraphics{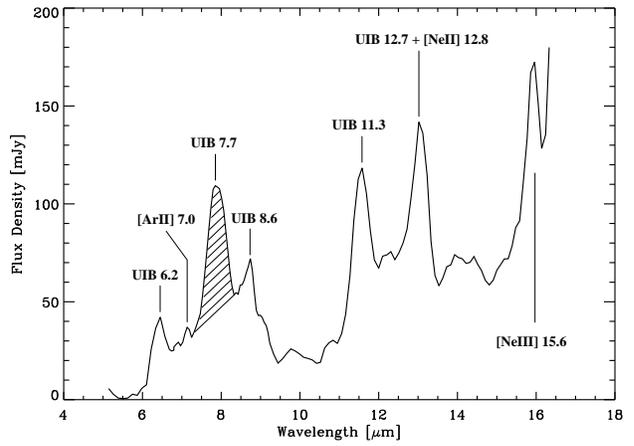}}
  \caption{ Mid-infrared spectrum of knot 2 in VV\,114W obtained with
  an aperture of 5{\arcsec} in diameter, centered on the peak of
  emission located near the bright \ion{H}{ii} region southeast of the
  nucleus of VV\,114W ($\alpha = 01^h07^m46.7^s, \delta =
  -17\degr30'27''$, J2000).  The flux of the UIB feature at
  7.7\,$\mu$m used to create the UIB map presented in Figure 2b was
  measured integrating the hatched area as illustrated on the figure.
  A similar technique was applied to the other features at 6.2 and
  11.3\,$\mu$m.  }
\label{fig:spec_vv114w}
\end{figure}

The spectral shape of the MIR emission throughout the extent of
VV\,114W was found rather constant across the entire disk as well as
on the overlap region towards the eastern component, varying only in
intensity. In Figure~\ref{fig:spec_vv114w} we present the spectrum of
a 5{\arcsec}-diameter aperture centered on the brightest peak of
emission in VV\,114W. As mentioned earlier, this is the location of
the so-called knot 2, which has the widest H$\alpha$ equivalent width
of the galaxy. The MIR spectrum is typical of what is observed in
other galaxies which are actively forming stars \citep[i.e.][ and
references therein ]{laurent}. As it is known, the spectrum can be
decomposed mainly into the contribution of the Unidentified Infrared
Bands (UIBs) clearly detected at 6.2, 7.7, 8.6, 11.3 and 12.7\,$\mu$m,
and that of a steeply rising continuum longward of 10\,$\mu$m.  The
UIBs are generally attributed to C=C and C\---H vibrations in
Polycyclic Aromatic Hydrocarbon molecules \citep[PAHs,
see][]{leger,li} which are stochastically heated by UV and optical
radiation, and since they are typically found in PDRs they are often
used to trace quiescent or more active star formation
\citep{helou}. The rising continuum on the other hand, is the most
prominent feature detected in MIR spectra of galactic \ion{H}{ii}
regions and galaxies harboring intense star forming activity. It is
thought to be produced by small dust particles with radius less than
10~nm \citep[Very Small Grains - VSGs - see][]{desert}, which are
heated up to temperatures $\sim$ 200--1000\,K by the strong radiation
field from hot young stars, and therefore it is considered as a good
tracer of massive star formation activity.  The presence of this
continuum in our spectrum -- though as we will see rather weak
compared to VV\,114E -- confirms that vigorous star formation is
underway in the overlap region of the merger, where the streaming
motions and shocking of gas clouds due to the interaction are probably
the strongest. This is in contrast to what one observed in the MIR
spectra of quiescent spiral galaxies where the contribution from the
VSGs is generally much lower \citep{roussel}. The overall extinction
in the area, as measured by the relative strength of the 6.2 and
11.3\,$\mu$m UIB features (see section~\ref{sec:nuc} for a description
of the method used), is small and the ISOCAM star formation activity
indicator as traced by the LW3/LW2 color ratio is just above $\sim$
1.4 \citep[see][]{laurent}.

Several ionic emission lines are also visible in the spectrum. At
7\,$\mu$m, we detect [ArII], which has a rather low ionization
potential (E$_p = 16~\mbox{eV}$) and is typically found in
starburst environments \citep{sturm}. Given our low spectral
resolution, the 12.8\,$\mu$m [NeII] line (E$_p = 22\,\mbox{eV}$) is
blended with the 12.7\,$\mu$m UIB feature and its strength can not be
accurately determined.  However, ISOCAM CVF observations of many
regions with moderate star forming activity and low extinction
\citep[i.e.][]{boulanger} indicate that the 12.7 and 11.3\,$\mu$m
UIB features scale with one another.  The fact that the 12.7\,$\mu$m
emission in VV\,114W is fairly strong suggests that a considerable
fraction of the feature flux is due to the [NeII] line. This would be
expected given the high H$\alpha$ flux of knot 2, a clear evidence of
massive extranuclear star formation activity.  The [NeII] line has
also been clearly observed with the larger aperture of ISO-SWS
\citep{thornley}, which provides further confidence on the reliability
of our detection. We finally observe an apparent bright feature at
15.6\,$\mu$m, which we attribute to [NeIII].  The profile of that
feature as well as the fact that this line was also observed with
ISO-SWS, lead us to believe that our detection is reliable, even
though the obvious memory effects of the detector make a determination
of its flux problematic.

\subsubsection{The MIR spectrum of the nuclear region in VV\,114E}
\label{sec:nuc}

\begin{figure}[!ht]  
\resizebox{\hsize}{!}{\includegraphics{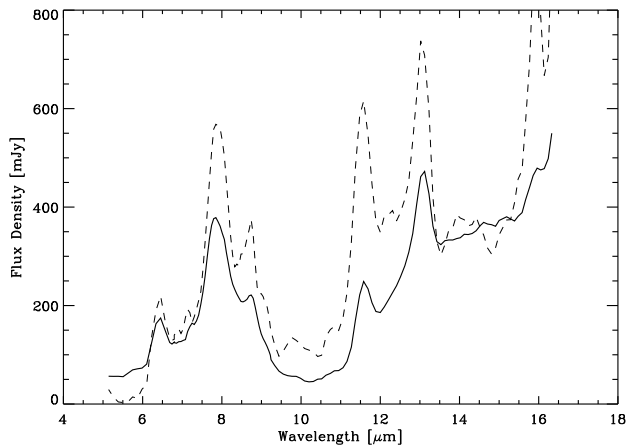}}
  \caption{The mid-infrared spectrum of the central region of VV\,114E
  using an aperture of 4.5{\arcsec}$\times$4.5{\arcsec}, taking
  into account the variable size of the PSF.  For comparison, we have
  reproduced the spectrum of VV\,114W ({\it dashed-line}) normalized
  to the rest frame 14--15\,$\mu$m continuum. Note the difference in
  the slope of spectra as well as the higher extinction in VV\,114E.}
  \label{fig:spec_vv114e}
\end{figure}

\begin{figure}[!ht]  
\resizebox{\hsize}{!}{\includegraphics{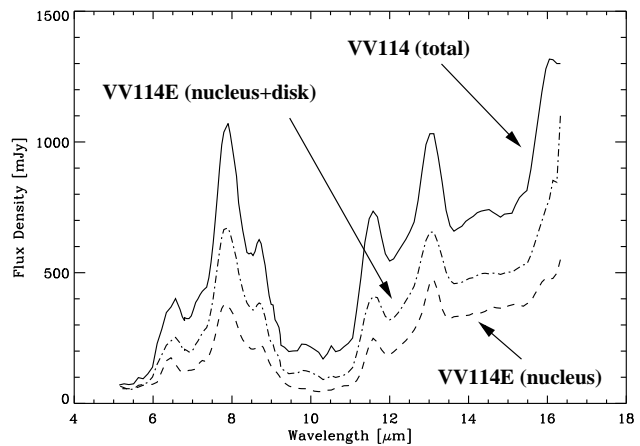}}
  \caption{The total MIR spectrum of VV\,114 (solid line) integrated
  over the full spatial extent of the merger. The 5--16\,$\mu$m
  SED of the nucleus of VV\,114E (dashed-line) and the spectrum of the
  whole eastern galaxy (disk+nucleus, dashed-dotted line) are also
  reproduced to show their relative contribution to the total
  emission.} \label{fig:spec_total}
\end{figure}

As discussed in section~\ref{sec:photo}, the MIR emission from the
central region of the eastern galaxy is dominated by the contribution
of its nucleus over the full wavelength range of the CVF.  In
Figure~\ref{fig:spec_vv114e} we present the spectrum of the central
4.5{\arcsec}$\times$4.5{\arcsec} (1.7\,kpc$\times$1.7\,kpc) region of
VV\,114E, after correcting it via a scaling factor which takes into
account the extended flux of the PSF lying outside the aperture. 
As in VV\,114W, we clearly detect the
presence of the UIB features as well as the [NeII] and [NeIII]
lines. However, comparing the spectra of the two galaxies, we note
that in VV\,114E the relative strength of the 8.6 and 11.3\,$\mu$m
features compared to the UIBs at 6.2, 7.7 and 12.7\,$\mu$m is much
lower, which implies a higher silicate absorption at 9.7\,$\mu$m in
the eastern galaxy.  The latter is also suggested by the 15\,$\mu$m to
9.7\,$\mu$m flux ratio, which for VV\,114E is $\sim 8$, nearly 4 times
that of normal galaxies \citep[see ][]{xu}.  The increased extinction
towards VV\,114E is also apparent as the fraction of the total flux
originating from VV\,114E reaches 65\% of the total shortward of
9\,$\mu$m and longward of 12\,$\mu$m, but falls to only 50\% near the
silicate band (see Table~\ref{tab:fluxes} and
Figure~\ref{fig:spec_total}).
To estimate the extinction in the central region of VV\,114E, we
reproduced the relative strength of the UIB features at 6.2 and
11.3\,$\mu$m using the observed spectrum of the M51 disk extinguished
by a screen model with different extinction laws.
This approach is justified since in
unobscured star forming regions, UIB features in the 5--13\,$\mu$m
range are known to present canonical spectral properties between one
another \citep{Dale01,roussel}.  The various extinction laws were
chosen to explore the effects of a variable strength in the
9.7\,$\mu$m silicate band, and they also differ in their level of
extinction between 4 and 8\,$\mu$m.  The results obtained are
summarized in Table~\ref{tab:extinction}.
In spite of the apparent variations between the different 
measurements, all values of $\tau$ are lower than 1, which is
consistent with an optically thin medium and making the screen
model a raisonable assumption.

\begin{table}[htbp]
\caption{Extinction in the central region of VV\,114E.}
\begin{tabular}{p{5cm}cc}
\hline 
\hline 
Extinction law & A$_v$ & $\tau$ at 10\,$\mu$m \\
\hline 
Draine (1989)                  & 4 mag & 0.33 \\
Mathis (1990)                  & 11 mag & 0.60 \\
Dudley \& Wynn-Williams (1997) & 5 mag & 0.28 \\
Lutz (1999)                    & 6 mag & 0.83 \\
Li \& Draine (2001)            & 5 mag & 0.41 \\
\hline
\hline
\end{tabular} \\
\vspace{.015cm} \\
\label{tab:extinction}
\end{table}

We note that we measure 
a lower extinction than \citet{soifer01} who derived
$\tau \sim 0.7$ for the southwestern nucleus of VV\,114E and $\tau
\sim$ 2--3 for the northeastern nucleus using the extinction law of
\citet{li} at 10\,$\mu$m. Our lower measurements are expected since
our larger ISOCAM beam samples a more diffuse optically thin dust
component surrounding the two nuclei of VV\,114E.

More intriguing is that VV\,114E displays a rather significant
continuum at short MIR wavelengths (5--6\,$\mu$m).  This ``bump'' is
absent in pure starburst spectra, but has already been observed
between 3--6\,$\mu$m in the SEDs of active galactic nuclei
\citep{lutz,mirabel99,laurent,lefloch} and it has been proposed as a
MIR diagnostic for an AGN \citep{laurent}. According to the unified
scheme of AGNs, such a continuum is thought to be produced by very hot
dust particles heated to nearly their sublimation temperature
(T\,$\sim$\,1000\,$\mbox{K}$ for silicate and
T\,$\sim$\,1500\,$\mbox{K}$ for graphite) within the torus of
molecular gas surrounding the nucleus \citep{krolik}.  The elevated
continuum at 5\,$\mu$m in our spectrum would be consistent with
emission from an enshrouded AGN. One should note however, that no
other telltale signs of an active nucleus such as the [NeVI] and [NeV]
lines have been detected on that galaxy, despite the rather modest
nuclear extinction of the nucleus. Could it be that this ``bump'' is
due to an ultracompact starburst instead? We discuss this possibility
in the following section.

\section{Discussion}
\label{sec:discuss}

\subsection{VV\,114~: the early stage of a future ULIRG}
\label{sec:sequence}

As we mentioned earlier, ULIRGs are formed when gravitational
instabilities occuring during galactic collisions, force large
concentrations of gas to the central regions of the merging disks,
fueling circumnuclear starbursts or enshrouded active nuclei
\citep[see][]{mihos96}. Observations of ULIRGs in the near and
mid-infrared have revealed that a large fraction (30--100\%) of the
infrared energy emitted by these objects is often contained in very
compact nuclear regions with linear scales of a few 100\,pc
\citep{scoville, soifer00,soifer01}.  Moreover, \citet{gao} have
shown a strong correlation between the CO luminosity and the nuclear
separation of merging galaxies, as well as an anti-correlation between
that distance of the nuclei and the L$_{\rm IR}$/M$_{\rm H_2}$ ratio.
In this context, one would expect that the ultraluminous phase would
appear as the final result of the merging, once the two progenitors
have nearly formed a single object.  However, it has been recently
reported that a population of ULIRGs may consist of widely separated
($\gtapp$ 10~kpc) pairs of galaxies in a rather early phase of
interaction \citep{murphy,dinh,meusinger}. This finding implies that
our interpretation on the mechanisms stimulating star formation during
interactions needs to be re-examined. More specifically,
\citet{murphy} have suggested that galaxies which have sufficiently
large reservoirs of molecular gas and experience more than one close
encounter before their final merger, may undergo multiple stages of
ultraluminous activity (L$_{\rm IR}\gtapp10^{12}$\,L$_{\sun}$)
triggered at each encounter of their disks while remaining in the LIRG
stage during the in-between periods.

Several observational evidence lead us to believe that VV\,114 is in
fact in such a ``quiescent LIRG'' phase. The first one is the moderate
separation ($\sim$ 6~kpc in projection) between the two interacting
galaxies, and the very extended molecular gas distribution as revealed
by the CO observations. The gas kinematics are also characterized by
large velocity gradients and non-rotational signatures dominated by
inflow/outflow radial motions, and even the distribution of the
stellar light in the central regions has not yet settled into an
$r^{1/4}$ profile \citep{yun,scoville}. Yet, despite the early stage
of its evolution, VV\,114 is already one of the brightest infrared
objects in the local Universe \citep{soifer87}, and harbors compact
nuclear and extranuclear starbursts with strong MIR emission.  We note
that such spots of star formation in the overlap region of interacting
galaxies are not uncommon when the progenitors are late type galaxies
\citep{mirabel98,gallais,xu}. Another characteristic of VV\,114,
yet unusual in LIRGs, is the similar curves-of-growth at 2.2 and
12.5\,$\mu$m in the central 1\,kpc region of VV\,114E
\citep{soifer01}. What could be the cause for this behavior? A
possible explanation might be related to the fact that even though the
system is highly IR luminous, it is dynamically young with its two gas
rich components still violently interacting. This would imply that not
only the old stellar population but large quantities of dust grains
coupled to the gas originally associated with the disks of VV\,114E
and VV\,114W before their interaction, have not yet relaxed and they
are distributed around the galaxies following loops and tails in
the gaseous component. The strong ionization field resulting mainly
from the massive starbursts in VV\,114E as well as in the overlap
region, would then permeate this extended volume of gas and dust, and
could cause the dust grains of smaller size to be heated sufficiently
in order to emit in the thermal MIR.

The large-scale spatial coincidence between the CO and MIR emission
shows that the conversion of molecular gas into stars is still
ongoing, not only in compact regions where the gravitation potential
is responsible for fueling the starburst with gas, but also in more
extended areas where shocks and density enhancements induced by the
dynamical perturbations of the collision may also trigger a star
forming activity. Taking into account the estimated age of the
collision \citep[$\sim$ 3--4\,$\times$\,10$^8$ yrs,][]{yun} and 
the huge reservoir
of gas still available (5.1$\times$10$^{10}$ M$_{\sun}$), it is very
likely that the star formation process will continue for a least a few
10$^8$\,yrs before the two galaxies finally coalesce. As suggested by
\citet{yun95} to illustrate how the gas density and IR luminosity
efficiency anti-correlate with nuclear separation in merging systems,
VV\,114 could evolve into a more advanced merger like Mrk\,273, to
finally terminate in a ULIRG similar to Arp\,220.

\subsection{Is there a buried AGN in VV\,114E~?}
\label{subsec:agn}

The question whether AGN or compact starbursts are responsible for the
production of the bulk of the luminosity of LIRGs/ULIRGs is a
pertinent one. The fact that both phenomena usually co-exist makes
quantifying their contribution rather problematic \citep[see][ for a
discussion]{laurent}. Recently, large scale surveys conducted at MIR
wavelengths with ISO \citep{lutz} have provided evidence that roughly
70\% of the energy budget in the Universe would originate from star
forming activity, though it has been suggested from submillimeter
surveys that the contribution from AGNs could be actually more
important \citep{mcmahon}.

The colors of VV\,114E are very red indicating a significant
absorption of UV and optical light by large concentrations of dust and thus
one may wonder whether or not an active nucleus, hidden in the
visible, could be lurking in it.  Apriori though, several observational
evidence argue against it. Not only the cold IRAS color of VV\,114
($f_{25 \mu m}/f_{60 \mu m}$ = 0.18) indicates a predominant star
formation activity, but also the two nuclei of VV\,114E show extended
emission at both NIR \citep{knop,doyon} and radio \citep{condon91}
wavelengths, which suggests that the observed continuum has a stellar
origin.  Moreover, the NIR spectroscopy of VV\,114E indicates that the
gas is photo-ionized by young OB stars rather than an AGN-like engine
\citep{doyon}, which reinforces the idea that the bulk of the infrared
emission from the galaxy may be attributed to an episode a vigorous
star formation.

However, using a J--H vs H--K color diagram, it can be shown that the
very central region of VV\,114E exhibits near-infrared colors similar
to that observed in warm LIRGs \citep{scoville}. Furthermore, models
fitting the NIR SED of VV\,114E indicate that 50\% of the K-band
continuum observed in the two nuclei originates from hot dust emission
heated to temperature $\sim$~800\,K \citep{doyon}. Unfortunately, even
the radio data of \citet{condon91} are not conclusive {\em for the two
nuclei} of VV\,114E as no high resolution radio spectral indexes which
could be used to identify a synchrotron or a flat thermal origin of
their emission are available. Finally, as it was shown by
\citet{soifer01} as well as from ISOCAM data, only of a fraction of 
the MIR emission seen by IRAS is associated with the nuclear region of
VV\,114. Consequently the IRAS colors do not characterize the nuclear
activity in this galaxy and can not be used to exclude a
dust-enshrouded active nucleus in VV\,114E.

In fact, our MIR spectra of VV\,114E do reveal a strong emission at
5\,$\mu$m, and a rather flat 5--16\,$\mu$m continuum compared to
typical template ISOCAM spectra of \ion{H}{ii} regions
\citep{laurent}. The specific signatures of PDRs, \ion{H}{ii} regions
and AGNs in the mid-infrared have been thoroughly studied by
\citet{genzel} and \citet{laurent} who, using the UIB to continuum
flux ratio of MIR templates to characterize each of the different
types of activity, have both developed diagnostics to distinguish
between starbursts and active nuclei in galaxies. Up to now, the type
of MIR spectrum seen from VV\,114E has only been found in AGNs and it
is thus considered as a solid indicator for the presence of an active
nucleus which is dominant in the MIR. Using the classification scheme
of \citet{laurent} it appears that an AGN could be contributing up to
40\% of the MIR flux in the eastern galaxy of VV\,114. Such a result
supports the strength of MIR observations to unveil dusty and hidden
AGNs which can not be detected at other wavelengths.

Recently, it has been proposed to combine the 3.3\,$\mu$m PAH emission
with the 3.4\,$\mu$m carbonaceous dust absorption to distinguish
between AGN and starburst activity in luminous galaxies
\citep{imanishi00,imanishi01}.  This technique could be particularly
powerful in detecting buried active nuclei in LINER- or
starburst-classified LIRGs/ULIRGs. It would be of great benefit if
this new diagnostic could be applied to VV\,114E so that we obtain a
better view on the 2--5\,$\mu$m window properties of that
merger. Establishing the link between what is observed in the K and L
band and the short (4--6\,$\mu$m) MIR wavelengths, which were observed
by ISO and will also be covered by SIRTF in the near future, is
clearly essential in order to better constrain the origin of the
different contributions and processes at work in this system.

\subsection{Implications for high redshift surveys}

Characterizing the physical properties of sources in the distant
Universe is one of the key issues to discriminate between
various models of galaxy formation. It has become increasingly
accepted that understanding the properties of galaxies such as VV\,114
may provide a valuable insight for this goal.  As discussed earlier,
this system consists indeed of two galaxies for which the presence of
dust is the cause of their dramatically different SEDs. Most of the
bolometric luminosity originates from VV\,114E which also dominates
the FIR emission from the system but is completely obscured by dust in
the far- and near-UV \citep{goldader}. On the other hand, the western
galaxy, VV\,114W, is bright in the UV and visible and contributes
only modestly to the global infrared emission. These ``antagonistic''
features have important consequences for the detection as well as the
apparent morphology of high redshift galaxies.  If we were to place
VV\,114 at $z \geq 1$, the spatial resolution of the deep surveys at
optical wavelengths \citep[e.g. HDF,][]{williams96} would not allow us
to disentangle the two components involved. One therefore would have
to rely only on its integrated properties. \citet{goldader} have shown
that VV\,114 could be easily observed at $z \sim 1.5$ in the optical
window, unlike the more luminous ULIRGs such as Arp\,220 which are too
faint in the UV. Due to the strong UV emission of VV\,114W it would be
detected even at z $\sim$ 3 with color, luminosity and size closely
resembling those of Lyman-break galaxies. Yet, the eastern component,
which as we have seen is the most energetically active, would be
invisible since its UV emission is completely absorbed and re-emitted
at longer wavelengths due to its high dust content
\citep[see Figure 3 and 4 of][]{goldader}. Obviously this will
provide a highly distorted and unrealistic view of the system since we
will be clearly unaware of the component which is responsible for the
bulk of its bolometric luminosity.  This would be similar to what
has been observed by \citet{Ivison} in the SCUBA source
SMMJ14011+0252, which is located at z=2.56 and harbors faint
unobscured companions within an otherwise large and opaque
star-forming system.

What evidence do we have though that VV\,114 is representative of the
high redshift galaxies?  Firstly we do know that mergers and
interactions were more frequent in the early Universe
\citep{lefevre}. Secondly the sources detected at 15\,$\mu$m in ISOCAM
deep MIR surveys \citep{aussel,elbaz} are mostly LIRGs rather than
ULIRGs \citep{elbaz02}. Finally, similarly to what is observed in
VV\,114, in most of the luminous infrared mergers at low-z the MIR
flux -- and possibly the FIR as well -- is found to originate mostly
from one component of the merging system
\citep{dinh,charmandaris02a,charmandaris02b}. The above observations
suggest that indeed galaxies such as VV\,114 could be very relevant in
probing galaxy evolution at high redshift.

\section{Conclusions} 
\label{sec:conclusions}

Using our ISOCAM MIR observations of the luminous infrared merger
VV\,114, we conclude that:

1) Nearly 65\% of the MIR emission detected in VV\,114 is associated
with its eastern component VV\,114E. Even though most of the flux from
VV\,114E seems to originate from the central 1\,kpc region, we also
observe a diffuse and extended component associated with the galactic
disk which is not detected in the ground-based MIR images of the
system. This is in stark contrast with other LIRGs/ULIRGs which
typically show a compact MIR emission. The global 5--16\,$\mu$m SED of
the eastern component suggests that most of its mid-infrared
luminosity is powered by vigorous starburst activity.

2) Almost 100\% of the IRAS 12\,$\mu$m flux of the merger has been
retrieved by ISOCAM and can be found within an area of 6\,kpc in
diameter.

3) The MIR emission from the western component VV\,114W is more
diffuse and does not peak at its nucleus. Rather, it displays a local
maximum near knot 2, a bright \ion{H}{ii} region located in the overlap
area of VV\,114E and VV\,114W.

4) The spectrum of VV\,114E displays an elevated continuum at
5\,$\mu$m which is typically observed in MIR spectra of active nuclei
as it is thought to originate from very hot dust ($\sim$\,1000-1500~K)
heated by the intense radiation field encountered in the vicinity of
AGNs. It could therefore reveal the presence of an AGN which, inconspicuous 
even at optical and radio wavelengths, contributes up to 40\%
of the MIR emission from the nucleus of VV\,114E.

5) At kpc scales, we note a generally good agreement between the
spatial distributions of the hot dust as seen in the MIR, the cold
dust observed at submillimeter wavelengths as well as the molecular
gas traced with the CO line. The displacement among their peaks
of emission may result from temperature gradients of the dust in the
overlap region of the two galaxies.

6) A LIRG similar to VV\,114 placed at high-z would be easily detected
in optical deep surveys. Yet, the derived properties, and especially
its estimated extinction, would be those of VV\,114W since the eastern
galaxy, which dominates the FIR/bolometric luminosity is obscured in the
UV and visible. Consequently, any correlation between the {\it
rest-frame} UV slope of its continuum and its bolometric luminosity
would be inaccurate as they would trace different components of the
system.

\begin{acknowledgements}  
We have greatly appreciated the assistance of R. Gastaud (Saclay) for
his help on the reduction and the analysis of the ISOCAM data. We
would like to thank D. Frayer (Caltech) for providing us with his SCUBA
maps and N. Scoville (Caltech) for making the reduced HST/NICMOS
images of VV\,114 publicly available. ELF wishes to express extensive
thanks to P.-A. Duc (Saclay) for fruitful discussions on numerous
aspects related to this work and his careful reading of the
manuscript. VC would like to acknowledge B.T. Soifer, G. Neugebauer
and J. Mazzarella (Caltech) for their advice on the comparison and
interpretation of ISO data with previous work as well as the support
of JPL contract 960803. IFM acknowledges partial support from CONICET,
Argentina. Finally, we warmly thank our referee, R.\,Maiolino, for
carefully reviewing the paper and particularly drawing our
attention on the origin of the bolometric luminosity of VV\,114.
Guest User, Canadian Astronomy Data Center, which is operated by the
Dominion Astrophysical Observatory for the National Resarch Council of
Canada's Herzberg Institute of Astrophysics.
\end{acknowledgements}

\appendix

\section{What is the IRAS 12\,$\mu$m flux of VV\,114?}

Since the analysis of our data revealed an extended MIR emission
surrounding VV\,114 we were very interested in comparing our
measurements with the IRAS 12\,$\mu$m flux of the galaxy. Our goal was
to accurately calculate what fraction of the IRAS flux was also
detected by ISOCAM. The two most commonly quoted values for the
12\,$\mu$m IRAS flux are 0.98$\pm$0.041 Jy, from the BGS sample
presented by \citet{soifer89}, and the more recent 0.6782$\pm$10\% Jy,
based on the Faint Source Catalogue of \citet{Moshir90}. Recently
though, \citet{soifer01} using the SCANPI utility provided by IPAC
re-measured the 12\,$\mu$m IRAS flux of the galaxy over an aperture of
2 arcmin in diameter and found that it was 1.1\,Jy.

Since the IRAS Si:As detectors used for the 12\,$\mu$m survey were
0.75$\times$4.5 arcmin in size, we felt that it was quite likely that
the measured emission from VV\,114 could be contaminated by another
companion galaxy. Indeed, observing the field around VV\,114 we
noticed that there were several sources in its proximity with one of
them, IC\,1622, being just 3.1 arcmin to the southwest, and having a
recession velocity of 6343\,km\,s$^{-1}$ just 300\,km\,s$^{-1}$ higher
than VV\,114 \citep[see Figure 107 of][]{Hibbard02}. IC\,1622 has
integrated B and V magnitudes which are just 1\,mag fainter than
VV\,114 and since its 1.4\,GHz radio continuum flux is 7.4\,mJy
\citep{Condon98} it should clearly form stars and emit in the MIR.
The possible MIR/FIR contribution of IC\,1622 was also noted
by \citet{condon90} due to the observed offset of the IRAS centroid
from the radio position of VV\,114.

We proceeded by reexamine in detail the five individual IRAS scans
passing near VV\,114. All scans were running from the southeast to
northwest direction. We noticed that the profiles of the spectra for 4
of the scans display a ``bump'' approximately 2.5 arcmin from the
nominal position of VV\,114. The only scan which does not display this
feature is scan \#3 and this is the nearest scan to the galaxy,
reaching to a minimum angular distance of less than 0.5
arcmin. Integrating the line profile for only this scan we find that
the zero crossing flux value is 0.63\,Jy. The above evidence lead us
to believe that the ``real'' 12\,$\mu$m flux of the galaxy, as seen by
IRAS, is better represented by the point source profile fit to it as
measured by the Faint Source Catalogue and hence ISOCAM has indeed
detected, within our photometric uncertainties, nearly 100\% of the
IRAS flux.


\begin{thebibliography}{}

\bibitem[Arp(1966)]{arp}
   Arp, H. 1966, ApJS, 14, 1  

\bibitem[Aussel et al.(1999)]{aussel}  
   Aussel, H., Cesarsky, C., Elbaz, D., \& Starck, J. L. 1999, A\&A,
   342, 313

\bibitem[Boselli et al.(1997)]{boselli}
   Boselli, A., Lequeux, J., Contursi, A., et al. 1997, A\&A, 324, L13

\bibitem[Boulanger et al.(1996)]{boulanger}
   Boulanger, F., Reach, W.T., Abergel, A., et al. 1996, A\&A, 315, 325

\bibitem[Bryant \& Scoville(1999)]{bryant}
   Bryant, P., \& Scoville, N.~Z. 1999, AJ, 117, 2632

\bibitem[Cesarsky et al.(1996)]{cesarsky}  
   Cesarsky, C., Abergel, A., Agnese, P., et al. 1996, A\&A, 315, L32

\bibitem[Charmandaris et al.(1999)]{charmandaris99}
   Charmandaris, V., Laurent, O., Mirabel, F., et al. 1999, A\&A, 341,
   69

\bibitem[Charmandaris et al.(2002a)]{charmandaris02a}
   Charmandaris, V., Stacey, G. J, \& Gull, G. 2002a, ApJ, 571, (in
   press, astro-ph/0201278)

\bibitem[Charmandaris et al.(2002b)]{charmandaris02b}
   Charmandaris, V., Laurent, O., Le Floc'h, E., et al. 2002b, A\&A
   (submitted)

\bibitem[Chary \& Elbaz(2001)]{chary}
   Chary, R., \& Elbaz, D. 2001, ApJ, 556, 562

\bibitem[Combes(2001)]{combes}
   Combes, F. 2001, Fueling the AGN. In Lectures on the Starburst-AGN
   Connection, INAOE, ed. D. Kunth, I. Aretxaga (astro-ph/0010570)

\bibitem[Condon et al.(1998)]{Condon98}
   Condon, J. J., Cotton, W. D., Greisen, E. W., et al. 1998, AJ, 115,
   1693
	
\bibitem[Condon et al.(1990)]{condon90}
   Condon, J. J., Helou, G., Sanders, D. B., \& Soifer, B. T. 1990,
   ApJS, 73, 359

\bibitem[Condon et al.(1991)]{condon91}
   Condon, J. J., Huang, Z.-P., Yin, Q. F., \& Thuan, T. X. 1991, ApJ,
   378, 65

\bibitem[Coulais \& Abergel(2000)]{coulais}
   Coulais, A., \& Abergel, A. 2000, A\&AS, 141, 533

\bibitem[Dale et al.(2001)]{Dale01}
   Dale, D. A., Helou, G., Contursi, A., Silbermann, N. A., \&
   Kolhatkar, S. 2001, ApJ, 549, 215

\bibitem[D\'esert et al.(1990)]{desert}  
   D\'esert, F.-X., Boulanger, G., \& Puget, J.-L. 1990, A\&A, 237,
   215

\bibitem[Dinh et al.(2001)]{dinh}
   Dinh, V. T., Lo, K. Y., Kim, D. C., Gao, Y., \& Gruendl,
   R. A. 2001, ApJ, 556, 141

\bibitem[Doyon et al.(1995)]{doyon}
   Doyon, R., Nadeau, D., Joseph, R. D., et al. 1995, ApJ, 450, 111

\bibitem[Draine(1989)]{draine}
   Draine, B. T. 1989, in Proc. 22nd ESLAB Symp. on IR Spectroscopy in
   Astronomy, ed. B.H. Kaldeich, ESA SP-290, 93

\bibitem[Dudley \& Wynn-Williams(1997)]{dudley}  
   Dudley, C. C. \& Wynn-Williams, C. G. 1997, ApJ, 488, 720

\bibitem[Elbaz et al.(1999)]{elbaz}
   Elbaz, D., Cesarsky, C., Fadda, D., et al. 1999, A\&A, 351, L37

\bibitem[Elbaz et al.(2002)]{elbaz02}
   Elbaz, D., Cesarsky, C., Chanial, P., et al. 2002, A\&A, 384, 848

\bibitem[Frayer et al.(1999)]{frayer}
   Frayer, D. T., Ivison, R. J., Smail, I., Yun, M. S., \& Armus L.
   1999, AJ, 118, 139

\bibitem[Gallais et al.(1999)]{gallais}
   Gallais, P., Laurent, O., Charmandaris, V., et al. 1999, in The
   Universe as seen by ISO, ESA~SP-427, p.\,880

\bibitem[Gao \& Solomon(1999)]{gao}
   Gao, Y., \& Solomon, P. M. 1999, ApJ, 512, L99

\bibitem[Genzel et al.(1998)]{genzel}  
   Genzel, R., Lutz, D., Sturm, E., et al. 1998, ApJ, 498, 579

\bibitem[Goldader et al.(2002)]{goldader}
   Goldader, J. D., Meurer, G., Heckman, T. M., et al. 2002, ApJ,
   568, 651

\bibitem[Helou et al.(2000)]{helou}
   Helou, G., Lu, N., Werner, M., Malhotra, S., \& Silbermann, N.
   2000, ApJ, 532, L21

\bibitem[Hibbard et al.(2002)]{Hibbard02}
   Hibbard, J. E., Rupen, M. P., \& van Gorkom, J. H. 2002, in ``Gas
   and Galaxy Evolution'', ASP Conf. Ser. Vol. 240, p.\,659, Hibbard
   J.E., Rupen M.P., van Gorkom J.H. (eds)

\bibitem[Hughes et al.(1998)]{hughes}
   Hughes, D. H., Serjeant, S., Dunlop, J., et al. 1998, Nature, 394,
   241

\bibitem[Imanishi \& Dudley(2000)]{imanishi00}
   Imanishi, M., \& Dudley, C. C. 2000, ApJ, 545, 701

\bibitem[Imanishi(2001)]{imanishi01}
   Imanishi, M. 2002, ApJ, 569, 44

\bibitem[Ivison et al.(2001)]{Ivison}
   Ivison, R. J., Smail, I., Frayer, D., Kneib, J.-P., \& Blain, A.
   2001, ApJ, 561, L45

\bibitem[Jarrett et al.(1999)]{jarrett}
   Jarrett, T. H., Helou, G., Van Buren, D., Valjavec, E., \& Condon
   J. J. 1999, AJ, 118, 2132

\bibitem[Kessler et al.(1996)]{kessler}  
   Kessler, M. F., Steinz, J. A., Anderegg, M.E., et al. 1996, A\&A,
   315, L27

\bibitem[Knop et al.(1994)]{knop}
   Knop, R. A., Soifer, B. T., Graham, J. R., et al. 1994, AJ, 107,
   920

\bibitem[Krolik(1999)]{krolik}  
   Krolik, J. 1999, in Active Galactic Nuclei, Princeton Series in
   Astrophysics, Princeton University Press

\bibitem[Laurent(1999)]{laurent99}
   Laurent, O. 1999, Ph.D. Thesis, University of Paris~XI, France

\bibitem[Laurent et al.(1999)]{laurent99b}  
   Laurent, O., Mirabel, I. F., Charmandaris V., et al. 1999, in XIXth
   Moriond Astrophysics Meeting: Building the Galaxies: From the
   Primordial Universe to the Present, p.\,79 (astro-ph/0005377)

\bibitem[Laurent et al.(2000)]{laurent}
   Laurent, O., Mirabel, I. F., Charmandaris, V., et al. 2000, A\&A,
   359, 887

\bibitem[Le Floc'h et al.(2001)]{lefloch}
   Le Floc'h, E., Mirabel, I. F., Laurent, O., et al. 2001, A\&A, 367,
   487

\bibitem[Le F\`evre et al.(2000)]{lefevre}
   Le F\`evre, O., Abraham, R., Lilly, S. J., et al. 2000, MNRAS, 311,
   565

\bibitem[L\'eger \& Puget(1984)]{leger}  
   L\'eger, A., \& Puget, J.-L. 1984, A\&A, 137, L5

\bibitem[Li \& Draine(2001)]{li}
   Li, A., \& Draine, B. T. 2001, ApJ, 554, 778

\bibitem[Lutz et al.(1998)]{lutz}  
   Lutz, D., Spoon, H. W. W., Rigopoulou, D., Moorwood, A.F.M., \&
   Genzel R. 1998, A\&A, 505, L103

\bibitem[Lutz(1999)]{lutz99}
   Lutz, D. 1999, in The Universe as seen by ISO, ESA~SP-427, p.\,623

\bibitem[Mathis(1990)]{mathis}
   Mathis, J. S. 1990, ARA\&A, 28, 37

\bibitem[McMahon et al.(1999)]{mcmahon}
   McMahon, R. G., Priddey, R. S., Omont, A., Snellen, I., \&
   Withington, S. 1999, MNRAS, 309, 1

\bibitem[Meusinger et al.(2001)]{meusinger}
   Meusinger, H., Stecklum, B., Theis, C., \& Brunzendorf, J. 2001,
   A\&A, 379, 845

\bibitem[Mihos \& Hernquist(1996)]{mihos96}
   Mihos, J. C., \& Hernquist, L. 1996, ApJ, 464, 641

\bibitem[Mirabel et al.(1998)]{mirabel98}  
   Mirabel, I. F., Vigroux, L., Charmandaris, V., et al. 1998, A\&A,
   333, L1

\bibitem[Mirabel et al.(1999)]{mirabel99}  
   Mirabel, I. F., Laurent, O., Sanders, D. B., et al. 1999, A\&A,
   341, 667

\bibitem[Moshir et al.(1990)]{Moshir90}  
        Moshir., M., et. al.  1990, IRAS Faint Source Catalogue,
        ver. 2.0.

\bibitem[Murphy et al.(2001)]{murphy}
   Murphy, T. W, Soifer, B. T., Matthews, K., \& Armus, L. 2001, ApJ,
   559, 201

\bibitem[Okumura(1999)]{okumura}
   Okumura, K. 1999, in ISO beyond point sources: studies of extended
   infrared emission, ESA~SP-455, p.\,47

\bibitem[Ressler et al.(1994)]{ressler}
   Ressler, M. E., Werner, M. W., van, Cleve, J., \& Choa, H. 1994,
   Exp. Astron., 3, 277

\bibitem[Rieke \& Low(1972)]{rieke}
   Rieke, G. H., \& Low, F. J. 1972, ApJ, 176, L95

\bibitem[Rigopoulou et al.(1999)]{rigopoulou}
   Rigopoulou, D., Spoon, H. W. W., Genzel, R., et al. 1999, AJ, 118,
   2625

\bibitem[Roussel et al.(2001)]{roussel}
   Roussel, H., Sauvage, M., Vigroux, L., \& Bosma, A. 2001, A\&A,
   372, 427

\bibitem[Sanders \& Mirabel(1996)]{sanders96}
   Sanders, D. B., \& Mirabel, I. F. 1996, ARA\&A, 34, 749

\bibitem[Sanders et al.(1988)]{sanders88}
   Sanders, D. B., Soifer, B .T., \& Helias, J. H. 1988, ApJ, 325, 74

\bibitem[Scoville et al.(2000)]{scoville}
   Scoville, N.Z., Evans, A.S., Thompson, R., et al. 2000, AJ, 119, 991

\bibitem[Soifer et al.(1987)]{soifer87}  
   Soifer, B. T., Sanders, D. B., Madore, B. F., Neugebauer, G., \&
   Danielson, G. E. 1987, ApJ, 320, 238

\bibitem[Soifer et al.(1989)]{soifer89}
   Soifer, B. T., Boehmer, L., Neugebauer, G., \& Sanders, D. B. 1989,
   AJ, 98, 766

\bibitem[Soifer et al.(2000))]{soifer00}
   Soifer, B. T., Neugebauer, G., Matthews, K., et al. 2000, AJ, 119,
   509

\bibitem[Soifer et al.(2001)]{soifer01}
   Soifer, B. T., Neugebauer, G., Matthews, K., et al. 2001, AJ, 122,
   1213

\bibitem[Solomon et al.(1997)]{solomon}
   Solomon, P. M., Downes, D., Radford, S. J. E., \& Barrett,
   J. W. 1997, ApJ, 478, 144

\bibitem[Starck et al.(1999))]{starck} 
   Starck, J. L., Abergel, A., Aussel, H., et al. 1999, A\&AS, 134,
   135

\bibitem[Sturm et al.(2000)]{sturm}
   Sturm, E., Lutz, D., Tran, D., et al. 2000, A\&A, 358, 481

\bibitem[Thornley et al.(2000)]{thornley}
   Thornley, M. D., F\"orster Schreiber, N. M., Lutz, D., et al.
   2000, ApJ, 539, 641

\bibitem[Williams et al.(1996)]{williams96}
   Williams, R. E., Brett, B., Dickinson, M., et al. 1996, AJ, 112,
   1335

\bibitem[Wilson et al.(2000)]{wilson00}
   Wilson, C. D., Scoville, N. Z., Madden, S. C., \& Charmandaris, V.
   2000, ApJ, 542, 120

\bibitem[Xu et al.(2000)]{xu} 
   Xu, C., Gao, Y., \& Mazzarella, J., et al. 2000, ApJ, 541, 644

\bibitem[Yun et al.(1994)]{yun}
   Yun, M. S., Scoville, N. Z., \& Knop, R. A. 1994, ApJ, 430, L109

\bibitem[Yun \& Scoville(1995)]{yun95}
   Yun, M. S., \& Scoville, N.Z. 1995, ApJ, 451, L45

\end{thebibliography}
\end{document}